\documentclass[11pt,twoside]{article}

\usepackage{galev06}
\usepackage{epsfig}
\usepackage{psfig}
\usepackage{lscape}

\begin{document}

\title{Simulations of Deep Extra-galactic Surveys with Herschel-SPIRE}   

\author{T. J. Waskett, B. Sibthorpe, M. J. Griffin}  

\affil{Cardiff University, Queens Buildings, The Parade, Cardiff. CF24 3AA. UK} 
%%% Fill in author affiliations

\begin{abstract} %%% Abstract to run on from here. 

The SPIRE Photometer Simulator reproduces the entire Herschel-SPIRE system in a
modular IDL program.  Almost every aspect of the operation of SPIRE can be
investigated in a systematic way to ensure that observations are performed in
the most efficient way possible when Herschel flies.  This paper describes some
of the work done with the Simulator to help prepare for large observing
programs such as deep extra-galactic, high-redshift surveys. 

\end{abstract}

%%% MAIN BODY OF TEXT GOES HERE. CONSULT "INSTRUCTIONS FOR AUTHORS USING
%%% LATEX2E MARKUP", SECTIONS 2.3-2.6 FOR HELP WITH EQUATIONS, FIGURES,
%%% AND TABLES.

\section{Introduction}   %%% Top level section head (remove "%" symbol)

Expensive space missions with limited operating lifetimes, such as Herschel
\citep{pilbratt}, require efficient operation in order to obtain the best
possible value for money in terms of the quantity and quality of astronomical
data obtained.  Careful preparation of every aspect of the mission
needs to be undertaken, including how the various instruments on board the
spacecraft are operated to achieve optimum performance.  SPIRE \citep{matt1} is
one of three scientific instruments on board the Herschel spacecraft and covers
the longer wavelength range from $\sim~200 - 600~\mu$m with both a photometer
and an imaging Fourier Transform Spectrometer (FTS).  This paper describes how
the operation of the photometer half of the SPIRE instrument has been optimised
through the use of an instrument Simulator, and how the same Simulator can be
used to prepare for large observing programmes.

\subsection{The SPIRE Photometer Simulator} 

The Simulator is an IDL coded virtual version of the photometer half of the
Herschel-SPIRE system.  It incorporates as many of the physical instrumental
and telescope characteristics as it is possible to include in a computationally
practical and user-friendly program. Full details of the individual modules and
their interaction with each other are given in \citet{bruce1}.

Briefly, the user creates a suitable input sky -- as realistic or fantastic as
desired -- for each of the three SPIRE bands.  These are fed into the Simulator
where they are convolved with a representative beam profile and then 'observed'
with the bolometer detector arrays.  Parameters for the observation are
predefined by the user in the same way that a real observation would be
planned.  The astronomical power from the sky and the background radiation from
the telescope and internal instrument components are all passed into a module
containing a model of the individual detectors, which calculates their response
to the incident radiation.  This bolometer model also calculates and
superimposes realistic noise on the output detector time-lines.  The detector
time-lines are then low-pass filtered and sampled, in the same way as done by
the on-board electronics, to produce output voltage time-lines.  Additionally, a
pointing time-line is generated based on the observation parameters.

\section{Optimisation of Observing Modes}

SPIRE has several modes of operation \citep{matt1}, but the primary mapping
modes are scan and jiggle.  The SPIRE arrays are hexagonally close-packed,
feedhorn coupled bolometers, with a detector spacing of twice the beam full
width half maximum (FWHM), so they do not fully sample the sky in a single
pointing.  The two mapping modes account for this in different ways.  Jiggle
map mode is used to observe individual $4\arcmin \times 4\arcmin$ fields by
pointing the telescope in one direction and 'jiggling' the instrument's internal
beam steering mirror to fill the whole field area with data.  While large maps
can be created by tiling together individual jiggle map fields it is more
observationally efficient to use scan map mode, which involves scanning the
telescope across the sky at a constant rate to create a strip of data.  By
decelerating the telescope at the end of one strip, slewing perpendicular to
the scan direction and accelerating back up to scanning speed in the opposite
direction, a large map can be built up with successive, adjacent scan strips. 
The detector arrays are rotated slightly, relative to the scan direction, so
that the gaps between adjacent detectors are filled by other detectors
following on behind. 

Because of the nature of the bolometer detectors scan map data time-lines are
subject to $1/f$ noise -- excess noise at low frequencies, or drifts on long
time-scales -- that cause 'stripes' in the resultant maps (see left hand panel
of Figure~\ref{pic}. Jiggle map mode is immune to $1/f$ noise because the
signal is 'chopped', so the map is the difference between on- and off-source
signals.  \citet{bruce2} describe how the Simulator has been used to determine
the optimum parameters for operating SPIRE in scan map mode, principally how
the competing effects of the low-pass electronics filter and $1/f$ noise lead
to an optimum scan speed around $30\arcsec$s$^{-1}$.  Additionally, the angle
of rotation between the scan direction and the array orientation, and the
spacing between successive scan strips have been optimised to produce highly
uniform sky coverage.  

\section{Preparation for Deep Extra-Galactic Surveys}

With the scan map mode parameters optimised we can investigate how instrumental
characteristics will affect actual observations.  Part of the SPIRE Guaranteed
Time (GT) programme will involve deep extra-galactic, high-redshift surveys
that cover large areas of sky down to, and even below, the confusion limit.  To
see how $1/f$ noise is likely to affect the sensitivity of such surveys to
faint point sources we designed a series of 'observations' to be carried out
with the Simulator.  We used the GALICS model of galaxy evolution
\citep{galics} to produce input skies of a suitably realistic extra-galactic
cosmological field covering 1 sq. deg.  The field was populated with 58590
sources down to a $250~\mu$m band flux limit of 2 mJy (a confusion limit of 40
beams per source is $\sim 1000$ sources per sq. deg.)  Since the GALICS model
is based on a hierarchical galaxy formation model the clustering of the sources
is more realistic than a simple Poisson distribution, and so the confusion
noise should be closer to what will be experienced by SPIRE in flight.  

Two sets of simulated observations were performed, one with the $1/f$ noise
switched off, leaving simply white noise in the detector time-lines, and the
other with the $1/f$ noise switched on and at a level that matches the design
requirement of the SPIRE bolometers (a knee frequency of 100 mHz.)  This is a
worst case scenario for the real instrument as the majority of the detectors
exhibit less $1/f$ noise than the requirement.  The final sensitivity of
the combined simulations (for each noise test) was designed to approximate a
confusion limited observation.

A source detection routine was run on naive maps created from each noise test
to compare the relative sensitivity of the observations with and without $1/f$
noise.  The extracted source list was then cross-referenced with the input
source catalogue.  As expected, the map with the $1/f$ noise showed a
significantly higher detection threshold ($\sim 30\%$) and with a higher
fraction of spurious source detections.  Therefore, if left untreated, $1/f$
noise could seriously affect such a survey.  Integration times would need to be
$\sim 70\%$ longer to reach the desired detection threshold than expected from
the sensitivity estimates, which currently assume no contribution from $1/f$
noise.

However, there is much that can be done to alleviate $1/f$ noise, for example
iterative map making methods borrowed from CMB analysis, or filtering schemes. 
High-pass filtering involves removing all the low frequency modes from a
Fourier transform of the time-line data from each detector.  This eliminates
the long period drifts associated with $1/f$ noise, effectively removing the
stripes in scan maps.  Some point source flux is also lost in this process
however, so the cut-off frequency must be a compromise between this and
removing as much $1/f$ noise as possible.

We devised the following scheme to filter the data, based on common cleaning
routines used in radio astronomy: first construct a map from the unfiltered
data; detect the brightest sources and subtract a point source model for each
source from the time-series data; Fourier transform and high-pass filter the
data, using a cut-off frequency of half the $1/f$ knee frequency (50 mHz in
this case); inverse Fourier transform the data back into time-series and add
the bright point sources back on; finally, re-create a map from this filtered
data.  Removing the bright point sources before filtering prevents negative
dips appearing in the map either side of the sources in the scan direction. 
This reduces the likelihood of the filtering process adversely affecting
fainter source flux recovery.

\begin{figure}
\begin{center}
\epsfig{file=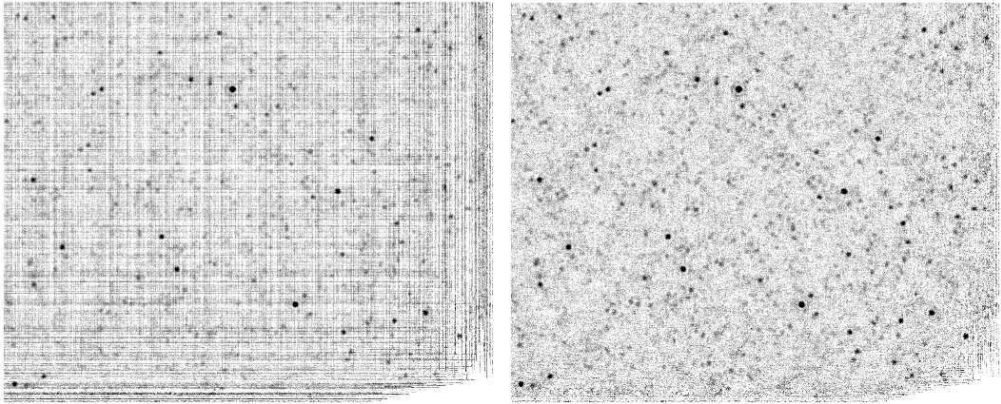,width=1.0\textwidth}

\caption{The lower right corner of the $250~\mu$m Simulator output map
(covering $\sim 35\arcmin \times 28\arcmin$) showing the effect of $1/f$ noise
(left) and how the simple filtering scheme improves the situation dramatically
(right).  Note how the SPIRE array has been scanned both horizontally and
vertically in this simulation to allow more sophisticated techniques, requiring
cross-linked data, to be tested.  The first Airy ring is also visible around
the brighter sources.}

\label{pic}
\end{center}
\end{figure}

Having made a new map we ran the source detection routine again.  The situation
was much improved with the detection threshold now only slightly higher than
the white noise case.  The number of spurious source detections was also much
reduced, although still higher than the white noise case.  Based on this
analysis we believe that the level of $1/f$ noise expected for SPIRE will not
cause a significant problem for these types of faint point source surveys. 
Simple filtering schemes can go a long way towards reducing $1/f$ noise effects
while more sophisticated methods may eliminate it entirely.  See
figure~\ref{pic} for a comparison of pre- and post-filtered images.

\section{Summary and Future Work}

The SPIRE photometer Simulator is a very powerful tool for investigating the
consequences of instrumental effects on the quality of astronomical data to be
obtained with SPIRE.  Not only has it helped in the optimisation of the
observing modes but it is also providing a head start in understanding the data
and how it will be best processed when the real thing becomes available.

As well as simulating cosmological surveys the Simulator is also being used to
simulate galactic (Sibthorpe, in preparation) and local Universe observations. 
Together, these simulations are central to the selection and development of the
map making code to be delivered as part of the SPIRE pipeline processing
package.  This process is currently underway and as launch approaches the
Simulator will be used to make further simulations to help refine the
performance of the map making algorithm as well as helping to develop optimised
source detection routines capable of fully exploiting the SPIRE data.

Finally, being a modular program the Simulator is fully customisable and can be
modified to represent any other facility of a similar design to SPIRE, or even
to help design the next generation of far-IR observatories.  

%\acknowledgements %%% Text of acknowledgements runs on after this command.

\end{document}